\documentclass[aps,prd,twocolumn,superscriptaddress,bibnotes,longbibliography,preprintnumbers,floatfix,10pt]{revtex4-2}

\usepackage{graphicx}
\usepackage{xcolor}
\usepackage{amsmath,amsfonts,amsthm,amssymb}
\usepackage[colorlinks]{hyperref}
\usepackage{mathtools}
\usepackage{bm}
\usepackage{multirow}
\graphicspath{{./figures/}}

\newcommand{\orcid}[1]{\href{https://orcid.org/#1}{\includegraphics[width=10pt]{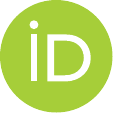}}}

\hypersetup{
    pdfnewwindow=true,
    colorlinks=true,
    linkcolor=violet,
    citecolor=violet,
    filecolor=violet,
    urlcolor=violet
}

\begin{document}
\preprint{FERMILAB-PUB-26-0451-T}

\title{Dark matter energy exchange in stars orbiting supermassive black holes}

\author{Stephan A. Meighen-Berger \orcid{0000-0001-6579-2000}\,}
\email{stephan-meighen-berger@uiowa.edu}
\affiliation{Department of Physics and Astronomy, \href{https://ror.org/036jqmy94}{University of Iowa}, Iowa City, IA 52242, USA}

\author{R. Andrew Gustafson \orcid{0000-0002-4794-7459}\,}
\email{gustafr@vt.edu}
\affiliation{Center for Neutrino Physics, Department of Physics,  \href{https://ror.org/02smfhw86}{Virginia Tech}, Blacksburg, Virginia 24061, USA}

\author{Nicole~F.~Bell \orcid{0000-0002-5805-9828}\,}
\email{n.bell@unimelb.edu.au}\affiliation{ARC Centre of Excellence for Dark Matter Particle Physics, \\
School of Physics,  \href{https://ror.org/01ej9dk98}{The University of Melbourne}, Victoria 3010, Australia}

\author{Jayden L. Newstead \orcid{0000-0002-8704-3550}\,}
\email{jayden.newstead@unimelb.edu.au}
\affiliation{ARC Centre of Excellence for Dark Matter Particle Physics, \\
School of Physics,  \href{https://ror.org/01ej9dk98}{The University of Melbourne}, Victoria 3010, Australia}

\author{Sandra Robles  \orcid{0000-0002-6046-8217}\,}
\email{srobles@fnal.gov}
\affiliation{Astrophysics Theory Department, Theory Division,  \href{https://ror.org/020hgte69}{Fermi National Accelerator Laboratory}, Batavia, Illinois 60510, USA}
\affiliation{Kavli Institute for Cosmological Physics, \href{https://ror.org/024mw5h28}{University of Chicago}, Chicago, Illinois 60637, USA}

\author{Ian M. Shoemaker \orcid{0000-0001-5434-3744}}
\email{shoemaker@vt.edu}
\affiliation{Center for Neutrino Physics, Department of Physics, \href{https://ror.org/02smfhw86}{Virginia Tech}, Blacksburg, Virginia 24061, USA}

\date{\today}

\begin{abstract}
Stars on tight orbits around the supermassive black hole at the Galactic Center pass through regions where the dark matter~(DM) density may be strongly enhanced. We compute the orbit-averaged DM-induced energy exchange for S4714 as an example. It is a star on an exceptionally close and relativistic orbit around Sagittarius~A*. For a spiked dark matter profile, the exchange reaches the stellar luminosity at $\sigma_{\chi p} \sim 10^{-36}~\mathrm{cm}^2$ for MeV--GeV masses and $\sigma_{\chi e} \sim 5\times10^{-38}~\mathrm{cm}^2$ for sub-MeV masses, opening a new annihilation-free route toward dark-star phases. These cross sections lie within the range predicted by freeze-in scenarios and are consistent with cosmic-ray--boosted and solar-reflection dark matter constraints.
\end{abstract}

 \maketitle


\section{Introduction}
\label{sec:intro}

\begin{figure}[t]
\centering
\includegraphics[width=0.95\columnwidth]{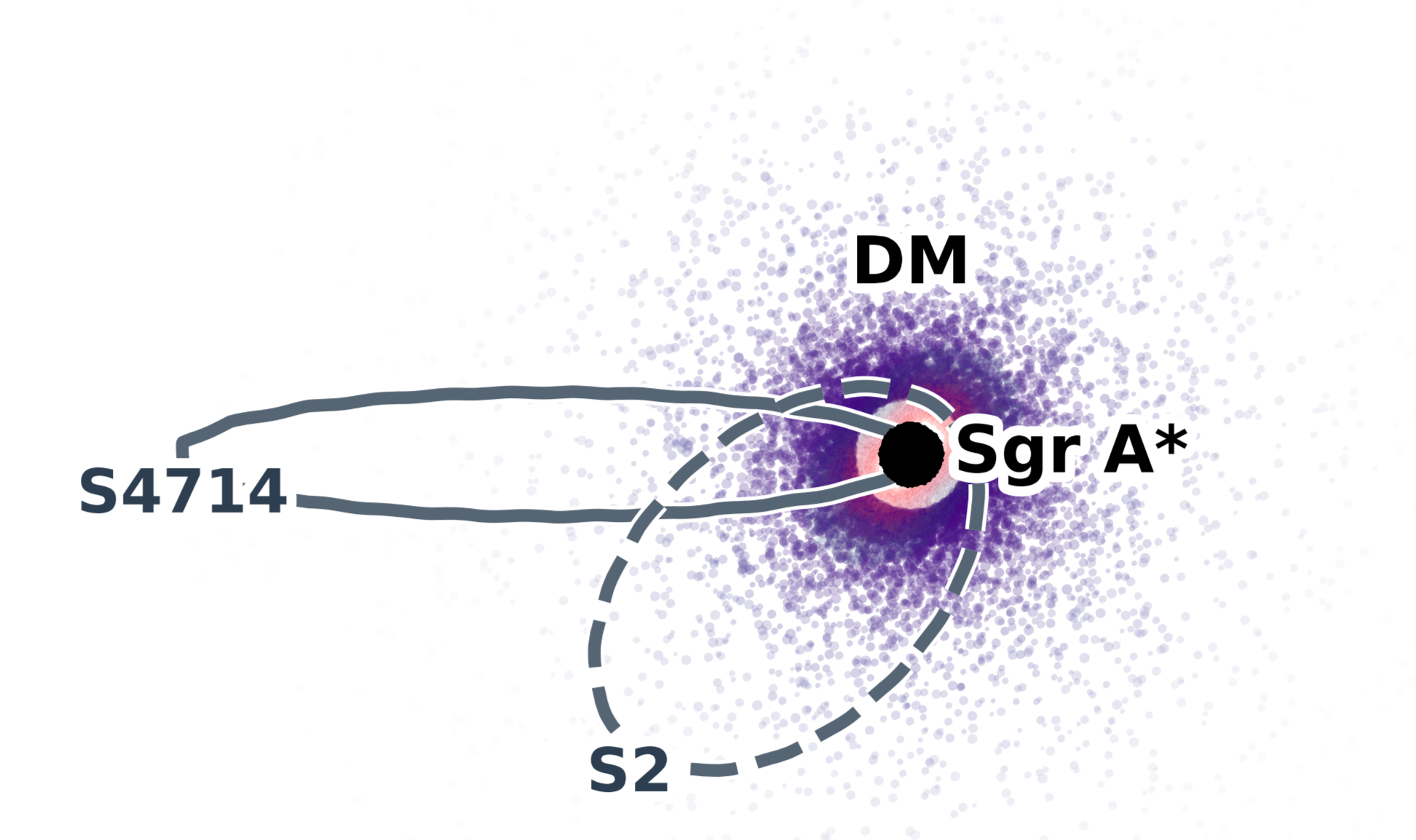}
\caption{Schematic illustration of a star on a highly eccentric orbit around Sagittarius~A* passing through a region of enhanced dark matter density. Elastic scattering between dark matter and stellar constituents (electrons and protons) can lead to a net transfer of energy, resulting in heating or cooling of the star depending on the DM mass and orbital velocity, with the net effect determined by integrating over the complete orbit.}
\label{fig:sketch}
\end{figure}

Sagittarius~A* (Sgr A*), the supermassive black hole (SMBH) at the center of the Milky Way, is a natural laboratory for probing gravity and particle physics under extreme conditions. Precision astrometry, particularly of the star S2, has measured the black hole's mass and distance and enabled tests of general relativity in the strong-field regime~\cite{Eisenhauer:2005cv, Martins:2007rv, GRAVITY:2018ofz}. The more recent discovery of an even more eccentric star S4714, with a pericenter distance and velocity of $\mathcal{O}(10~\mathrm{AU})$ and $\mathcal{O}(10^{-2}c)$ respectively~\cite{2020ApJ...899...50P} (see Table~\ref{tab:stars} for more details), makes it a uniquely sensitive probe of the DM density profile.
 
Dark matter spikes form via the adiabatic growth of the host SMBH, steepening the density profile and increasing the DM density by a factor of $10^8$--$10^9$ relative to an NFW profile~\cite{Gondolo:1999ef, Quinlan:1994ed,Sadeghian:2013laa}. The existence of a spike around Sagittarius~A* remains uncertain. Gravitational heating from stars, mergers, and dark matter self-annihilation can weaken or erase it~\cite{Bertone:2005hw, Bertone:2005xv, Balaji:2023hmy,Akita:2025dhg}. However, current observations do not rule it out~\cite{Lacroix:2018zmg, GRAVITY:2024tth, Shen:2023kkm, Chattopadhyay:2026kbm}.. S4714's exceptional orbit penetrates the innermost region of any such spike, making it a particularly sensitive probe.
 
Here we explore the elastic scattering of DM in the star, which continuously exchanges thermal energy with the stellar plasma, depositing or extracting energy depending on the DM mass and velocity. Averaged over an orbit, this can produce a significant net heating or cooling of the star. Previous work has considered DM capture and annihilation in stars~\cite{Scott:2008ns, John:2023knt} and orbital perturbations from elastic scattering~\cite{Acevedo:2025rqu,Gustafson:2025ypo}. Unlike capture-and-annihilation scenarios, the mechanism studied here does not require DM self-annihilation. It therefore applies to asymmetric DM and other models without annihilation.
 
Figure~\ref{fig:sketch} illustrates a star on a tight orbit around Sagittarius~A* passing through the dark matter distribution. We compute the orbit-averaged DM-induced energy exchange for S4714 as a fraction of stellar luminosity across a wide DM mass range, for DM--proton and DM--electron scattering and for NFW and spiked profiles. We refer to this energy transfer as a ``luminosity" as it has the same units of $\mathrm{GeV \, s^{-1}}$, although it does not directly correspond to an observation. We find that, for a spiked profile, elastic DM scattering can produce an orbit-averaged luminosity comparable to the stellar luminosity of S4714 at cross sections not yet excluded by experiment. The relevant cross sections lie within the range predicted by freeze-in scenarios~\cite{Elor:2021swj, Bhattiprolu:2022sdd}, and are independently accessible to cosmic-ray-boosted and solar-reflection DM searches.
 
The structure of the paper is as follows. Section~\ref{sec:densities} describes the DM density profiles and the energy-exchange mechanism. Section~\ref{sec:results} shows our results and discussion, with conclusions in Section~\ref{sec:conclusion}.


\section{Dark matter densities and Energy Exchange}
\label{sec:densities}

We adopt an NFW profile~\cite{Navarro:1995iw} for the dark matter distribution outside the spike region, with inner slope $\gamma$ and density
\begin{equation}
\rho_{\mathrm{NFW}}(r) = \frac{\rho_s}{(r/r_s)^\gamma(1+r/r_s)^{3-\gamma}},
\end{equation}
where $r_s = 20~\mathrm{kpc}$ is the scale radius for the Milky Way halo~\cite{Cirelli:2010xx} and $\rho_s$ is the scale density, fixed by requiring the local DM density at the Sun's galactocentric distance $r_\odot = 8.5~\mathrm{kpc}$ to match the measured value $\rho_\odot = 0.4~\mathrm{GeV~cm^{-3}}$, giving $\rho_s = \rho_\odot(r_\odot/r_s)^\gamma(1 + r_\odot/r_s)^{3-\gamma}$ (i.e.\ $\rho_s \approx 0.35~\mathrm{GeV~cm^{-3}}$ for $\gamma = 1$). We set $\gamma = 1$, though values in the range $0.5 \lesssim \gamma \lesssim 1.5$ are well motivated~\cite{Gnedin:2003rj, 2015MNRAS.448..713P, Iocco:2016itg, Hooper:2016ggc,
Baumgart:2025dov, Akita:2025dhg}.

\begin{figure}[t]
\centering
\includegraphics[width=0.95\columnwidth]{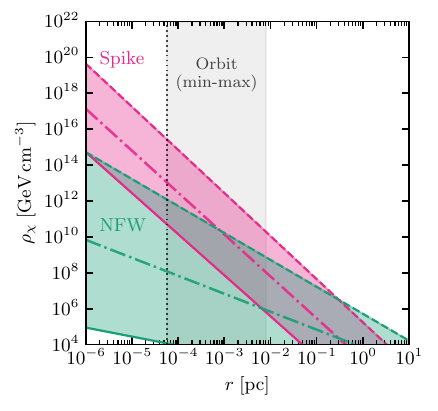}
\caption{DM density as a function of radius. The pink band shows the extreme spike profile combined with an NFW profile, as predicted by Ref.~\cite{Gondolo:1999ef} for adiabatic black-hole growth. The green band indicates the standard NFW profile~\cite{Navarro:1995iw} without a spike. The grey shaded band denotes the orbit of S4714 around Sagittarius~A*~\cite{Martins:2007rv, 2020ApJ...899...50P}. Each band indicates different $\gamma$ values, 0.5 (solid), 1 (dashed dotted), and 1.5 (dashed) for the corresponding profile.}

\label{fig:density}
\end{figure}

We use the adiabatic spike profile as our fiducial model. When the central black hole grows adiabatically within a pre-existing NFW halo with inner slope $\gamma$, gravitational focusing steepens the profile to a power-law spike with index $\gamma_{\mathrm{sp}} = (9-2\gamma)/(4-\gamma)$~\cite{Gondolo:1999ef}. For a typical NFW profile, $\gamma = 1 \rightarrow \gamma_{\rm sp} = 7/3$. The spike density is
\begin{equation}
\rho_{\mathrm{sp}}(r) = \rho_{\mathrm{NFW}}(R_{\mathrm{sp}})\left(\frac{R_{\mathrm{sp}}}{r}\right)^{\gamma_{\rm sp}},
\label{eq:spike_fct}
\end{equation}
matched continuously to the NFW profile at the spike radius $R_{\mathrm{sp}} = 0.3~\mathrm{pc}$~\cite{Gondolo:1999ef, Quinlan:1994ed, Sadeghian:2013laa}. We take $M_{\mathrm{BH}} = 4.3\times10^6~M_\odot$~\cite{GRAVITY:2018ofz} for the mass of Sagittarius~A*. The composite profile is $\rho_\chi(r) = \rho_{\mathrm{sp}}(r)$ for $r \leq R_{\mathrm{sp}}$ and $\rho_{\mathrm{NFW}}(r)$ for $r > R_{\mathrm{sp}}$.

Figure~\ref{fig:density} shows the DM profiles in our analysis. Stellar gravitational heating and other processes can soften or erase the spike~\cite{Bertone:2005hw, Bertone:2005xv, Balaji:2023hmy}, though recent modeling suggests it can survive~\cite{karydas2026survivaldarkmatterspikes}. We therefore also show results for a standard NFW profile. Appendix~\ref{app:feedback} shows that the relevant cross sections do not substantially exceed the baryonic feedback threshold.

The GRAVITY Collaboration recently constrained the enclosed mass within S2's orbit to $\lesssim 1200~M_\odot$ ($1\sigma$)~\cite{GRAVITY:2024tth}, limiting how steep the spike can be. For our fiducial spike profiles, the enclosed mass within
S2's orbit ($\sim 0.01~\mathrm{pc}$) is $\sim 0.14~M_\odot$ for $\gamma = 0.5$ and $\sim 24~M_\odot$ for $\gamma = 1.0$, both well within the observational bound. The steep $\gamma = 1.5$ ($\gamma_\mathrm{sp} = 2.4$) profile, however, implies $\sim 4400~M_\odot$ enclosed, roughly a factor of four above the limit, and is therefore in tension with current observations. We retain it as an upper bound. We compute the DM-induced energy exchange for stars traversing these profiles.


\subsection{Analytic approach}
\label{sec:fully_analytic}
 
\paragraph{Physical picture.}
We first establish the sign of energy transfer by comparing the DM kinetic energy with the thermal energy of stellar material. Consider a DM particle of mass $m_\chi$ with velocity $v_\star$ in the stellar rest frame. We assume DM is at rest in the Galactic frame. DM bound to Sgr\,A$^*$ has additional orbital velocity at periapsis, increasing $L_\chi \propto v_{\rm rel}^3$ and making our results conservative lower bounds. Appendix~\ref{app:dm_velocities} quantifies this correction. Upon scattering with a target particle (electron or proton) at thermal velocity $v_{\mathrm{th}} = \sqrt{3k_B T/m_{\mathrm{target}}}$, energy can flow in either direction.
 
For \textbf{heavy dark matter} ($m_\chi \gtrsim 0.4~\mathrm{MeV}$ for S4714), the DM kinetic energy in the stellar rest frame,
\begin{equation}
    E_{\mathrm{kin}} = \frac{1}{2} m_\chi v_\star^2,
    \label{eq:kinetic_energy}
\end{equation}
exceeds the thermal energy of stellar constituents, $E_{\mathrm{th}} \sim k_B T$. For S4714 orbiting Sagittarius~A* with pericenter velocity
$v_\star \sim 24\,000~\mathrm{km~s^{-1}}$, a DM particle with $m_\chi = 10~\mathrm{MeV}$ carries $E_{\mathrm{kin}} \sim 30~\mathrm{keV}$ while the stellar core temperature is only $T \sim 10^7~\mathrm{K}$ (corresponding to $k_B T \sim 1~\mathrm{keV}$). Since $E_{\mathrm{kin}} \gg k_B T$ in this regime, kinetic energy transfers from the DM to the stellar plasma, leading to net heating. Captured DM that subsequently evaporates still deposits net energy equal to the entry kinetic energy minus the escape energy.
 
For \textbf{light dark matter} ($m_\chi \lesssim 0.4~\mathrm{MeV}$ for S4714), the situation reverses. A DM particle with $m_\chi = 10~\mathrm{keV}$ carries only $E_{\mathrm{kin}} \sim 30~\mathrm{eV}$ in S4714's orbital environment, far less than the thermal energy of stellar plasma ($\sim 1~\mathrm{keV}$). DM--plasma scattering therefore transfers energy from the star to the DM, cooling it.
 
The transition between these regimes occurs when $E_{\mathrm{kin}} \sim E_{\mathrm{th}}$, i.e., when
\begin{equation}
    \frac{1}{2} m_\chi v_\star^2 \sim \frac{3}{2} k_B T,
    \label{eq:transition_condition}
\end{equation}
which gives the critical DM mass~\cite{Gould:1987ir, Gould:1987ju}
\begin{equation}
    m_\chi^{\mathrm{crit}} \sim \frac{3 k_B T}{v_\star^2}.
    \label{eq:critical_mass}
\end{equation}
For S4714 ($v_\star \sim 24,000~\mathrm{kms^{-1}}$, $k_B T \sim 1\mathrm{keV}$), this gives $m_\chi^{\mathrm{crit}} \sim 0.4~\mathrm{MeV}$.
 
\paragraph{Analytic luminosity.}
We now derive an analytic expression for the total energy exchange at periapsis. The net energy deposited per scatter is $\delta\varepsilon_{\mathrm{peri}} = \Delta E_{\mathrm{in}} - \Delta E_{\mathrm{out}}$. Here, each term follows from the transfer function
\begin{equation}
\Delta E(E_{\mathrm{ref}},\,m_{\mathrm{ref}}) = \frac{4m_p\left(1 + E_{\mathrm{ref}}/m_\mathrm{ref}\right)
E_{\mathrm{ref}}/m_\chi}{(1+m_p/m_\chi)^2 + 2E_{\mathrm{ref}}/m_\chi},
\label{eq:dE}
\end{equation}
with $\Delta E_{\mathrm{out}} = \Delta E(k_B T_{\mathrm{eff}},\,m_p)$ the energy a thermal proton transfers to the DM and $\Delta E_{\mathrm{in}} = \Delta E(\tfrac{1}{2}m_\chi v_{\mathrm{peri}}^2,\,m_\chi)$ the energy the DM deposits at periapsis. Here $m_{\mathrm{ref}}$ is the mass of the particle carrying $E_{\mathrm{ref}}$, and $(1 + E_{\mathrm{ref}}/m_{\mathrm{ref}})$ captures the leading relativistic correction. For DM--electron scattering, $m_p \to m_e$ throughout Eq.~(\ref{eq:dE}). The quantity $\delta\varepsilon_{\mathrm{peri}}$ changes sign at $m_\chi^{\mathrm{crit}}$, consistent with the physical picture above.
 
The instantaneous energy exchange rate at periapsis follows from multiplying the DM
transit rate by the scattering probability and the net energy per scatter,
\begin{equation}
    L_{\mathrm{peri}} = \pi R_\star^2\,\frac{\rho_\chi(r_{\mathrm{peri}})}{m_\chi}\,
    v_{\mathrm{peri}} \cdot P_{\mathrm{scat}}\,\delta\varepsilon_{\mathrm{peri}},
    \label{eq:lum_peri}
\end{equation}
where $P_{\mathrm{scat}} = R_\star n_p \sigma_{\chi p}$ is the mean number of scatters per transit. In the optically thin limit $P_{\mathrm{scat}} \ll 1$~\cite{Gould:1987ir}. Appendix~\ref{app:details} verifies that this holds for $\sigma_{\chi p} \lesssim 10^{-35}~\mathrm{cm}^2$.
 
The analytic luminosity, orbit-averaged via a scaling factor $f_{\mathrm{peri}}$, is then
\begin{equation}
    L_\chi = A \cdot f_{\mathrm{peri}} \cdot L_{\mathrm{peri}}(f_T),
    \label{eq:lum_fa}
\end{equation}
where $A \sim 0.14$ and $f_T \sim 0.4$ are correction factors from a $2\,M_\odot$ MESA stellar model~\cite{2011ApJS..192....3P,2013ApJS..208....4P,2015ApJS..220...15P,Paxton:2017eie,2019ApJS..243...10P,MESA:2022zpy} accounting for the non-uniform column density and profile-averaged temperature along each DM chord. Their definitions are in Appendix~\ref{app:fperi_deriv}. $f_{\mathrm{peri}} = 0.002$ follows from Eq.~(\ref{eq:f_peri}) of Appendix~\ref{app:fperi_deriv}.
 
Finally, the luminosity is bounded above by a geometric saturation rate at which every transiting DM particle deposits its full kinetic energy. At this point, increasing the cross section will not alter the heating rate. Here $\rho_{\mathrm{sp}}$ is in energy-density units (GeV~cm$^{-3}$), so $\rho_{\mathrm{sp}}/c^2$ is the mass density, giving
\begin{equation}
    L_{\mathrm{sat}} = \frac{\pi f_{\mathrm{peri}}}{2}\,\frac{R_\star^2}{c^2}\,\rho_{\mathrm{sp}}(r_{\mathrm{peri}})\,v_{\mathrm{peri}}^3. 
    \label{eq:lsat_fa}
\end{equation}
The full analytic luminosity is then
\begin{equation}
    L_\chi(\sigma_{\chi p}) = \min\!\left( A\,f_{\mathrm{peri}}\,L_{\mathrm{peri}}(f_T),\,
L_{\mathrm{sat}} \right).
\end{equation}
We also used the energy exchange formalism from~\cite{Gould:1987ir, Gould:1987ju} and found good agreement within $\sim 3\%$ compared to the formalism used here throughout the heating regime. Deviations near the transition reflect different effective-temperature definitions.
 
We now consider the scenario where $m_\chi = m_p$. The maximum elastic energy transfer per collision, $T_{\max}\sim v_{\mathrm{peri}}^2 m_\chi^2 m_p/(m_\chi+m_p)^2$, reduces to $v_{\mathrm{peri}}^2 m_\chi$, exactly canceling the $1/m_\chi$ in the DM flux. The deposited power at periapsis is then mass-independent,
\begin{equation}
L_{\mathrm{peri}}(m_\chi = m_p) \sim \pi R_\star^2\, v_{\mathrm{peri}}^3\, \rho_\chi(r_{\mathrm{peri}})\, P_{\mathrm{scat}}.
\end{equation}
Applying the correction factors  $A \sim 0.14$, $f_T \sim 0.4$, and $f_{\mathrm{peri}} \sim 2\times10^{-3}$ and for a spike profile at S4714's closest approach ($\rho_{\mathrm{sp}}(r_{\mathrm{peri}}) \sim 10^{13}~\mathrm{GeV\,cm^{-3}}$ for $\gamma_{\mathrm{sp}} = 7/3$),
\begin{equation}\label{eq:estimate}
L_{\chi}^{\mathrm{Spike}}(m_\chi = m_p) \sim P_{\mathrm{scat}} \times 3\times 10^{38}~\mathrm{GeV~s^{-1}}.
\end{equation}
Expanding $P_{\mathrm{scat}} = R_\star n_p \sigma_{\chi p}$, setting $R_\star = 2R_\odot$ and the column-averaged proton density $n_p \approx 3\times10^{23}~\mathrm{cm^{-3}}$,
\begin{equation}\label{eq:estimate_xsec}
L_{\chi}^{\mathrm{Spike}}(m_\chi = m_p) \sim \left(\frac{\sigma_{\chi p}}{10^{-36}~\mathrm{cm}^2}\right) \times 1\times 10^{37}~\mathrm{GeV~s^{-1}}.
\end{equation}
Setting this equal to $L_\star \approx 17.5\,L_\odot \approx 4.2\times10^{37}~\mathrm{GeV\,s^{-1}}$~\cite{2020ApJ...899...50P,2011ApJS..192....3P,2013ApJS..208....4P,2015ApJS..220...15P,Paxton:2017eie,2019ApJS..243...10P,MESA:2022zpy} gives $\sigma_{\chi p} \sim 4\times10^{-36}~\mathrm{cm}^2$. An NFW profile without the spike results in roughly four orders of magnitude lower heating.

\subsection{Monte Carlo method}
\label{sec:mc}
 
The Monte Carlo (MC) method simulates individual DM transits through the stellar interior using the full radial profiles $n_H(r)$ and $k_BT(r)$ from the same MESA $2\,M_\odot$ model. For each transit, the DM propagates along a chord through the star, accumulates optical depth, and undergoes elastic collisions with target velocities drawn from the local Maxwell--Boltzmann distribution. This naturally handles multiple scatters per transit when $P_{\mathrm{scat}} \gtrsim 1$, gravitational capture and evaporation of light DM ($m_\chi < m_{\mathrm{evap}}$), and the full velocity dependence of the energy transfer at each radial position. The orbit average uses Keplerian sampling. The mean energy deposited per transit $\langle\Delta E\rangle(m_\chi, \sigma_{\chi p}, v_{\mathrm{entry}})$ is evaluated from a pre-computed lookup table at each orbital phase and weighted by the local DM flux. Simulation details are in Appendix~\ref{app:mc_details}.

 
\section{Results and Discussion}
\label{sec:results}
 
We assess whether significant DM-induced heating or cooling of S4714 is possible, given prior constraints on DM scattering. Our results are presented for DM--proton and DM--electron scattering separately.
 
\subsection{DM--proton scattering}
 
\begin{figure}[t]
\centering
\includegraphics[width=0.95\columnwidth]{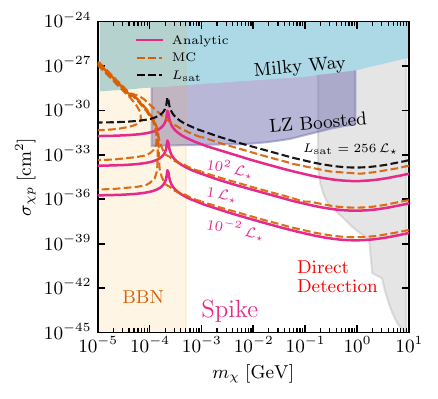}
\caption{Analytic orbit-averaged DM--proton energy exchange in S4714 as a fraction of stellar luminosity, for the $\gamma=1 \rightarrow \gamma_{\rm sp} = 7/3$ spiked profile across the full DM mass range. MC results are overlaid as a validation of the analytic approach. The two methods agree well across the heating regime. Discrepancies near the heating--cooling transition reflect the breakdown of the mean-scatter treatment discussed in Section~\ref{sec:discussion}. The experimental bounds are from LZ~\cite{LZ:2025iaw} CR-boosted DM search and the direct detection bounds adopted from Ref.~\cite{Billard:2021uyg} with data from CRESST~\cite{CRESST:2017ues, CRESST:2022dtl}, DAMIC~\cite{DAMIC:2020cut}, DarkSide~\cite{DarkSide:2018bpj, DarkSide:2018kuk}, and XENON1T\cite{XENON:2018voc, XENON:2019gfn}. The vertical shaded region is the BBN bound~\cite{Sabti:2019mhn, Krnjaic:2019dzc, Sabti:2021reh}. The Milky-Way constraint is from~\cite{Nadler:2019zrb}.}
\label{fig:p_mc_high}
\end{figure}
 
Figure~\ref{fig:p_mc_high} shows DM--proton energy exchange in S4714 as a fraction of stellar luminosity for the $\gamma = 1 \rightarrow \gamma_{\rm sp} = 7/3$ spiked profile. The exchange reaches the stellar luminosity at $\sigma_{\chi p} \sim 10^{-36}~\mathrm{cm}^2$ for $m_\chi \sim 1~\mathrm{GeV}$, near the current experimental limit. For sub-GeV masses, this contour lies below existing constraints. These results show that DM with cross sections below existing limits can significantly heat S4714.

\begin{figure}[t]
\centering
\includegraphics[width=0.87\columnwidth]{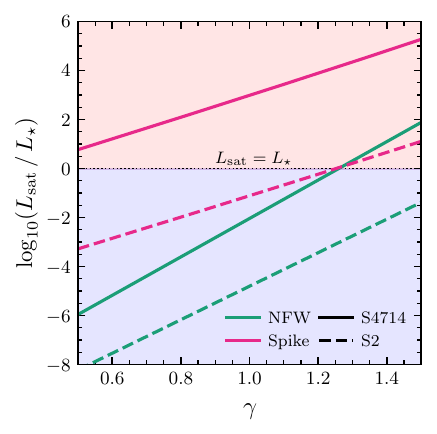}
\caption{Orbit-averaged saturation luminosity $L_{\mathrm{sat}}$ as a fraction of stellar luminosity for S4714 and S2, for NFW and Spike profiles with different $\gamma$. The strong dependence on $\gamma$ reflects the steep increase in periapsis density with inner slope.}
\label{fig:Saturation}
\end{figure}

The energy transfer is bounded above by $L_{\mathrm{sat}}$ (Eq.~\eqref{eq:lsat_fa}), which depends on the star's size, orbit, and DM density. At fixed geometry, $L_{\mathrm{sat}}$ depends only on the DM density profile. In Figure~\ref{fig:Saturation}, we show the saturation bound for a range of NFW DM profiles with and without spikes. We see that, for S4714, an NFW profile without a spike for $\gamma >  1.25$ can have a saturation luminosity exceeding $L_\star$.
 
\begin{figure}[t]
\centering
\includegraphics[width=0.95\columnwidth]{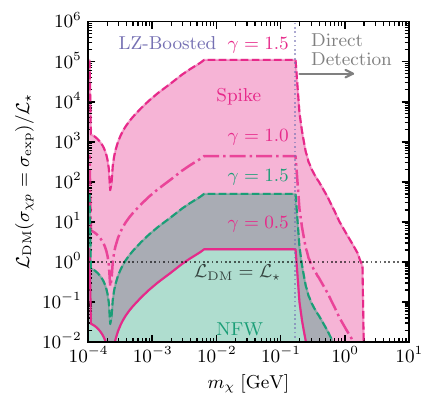}
\caption{Orbit-averaged DM-induced energy exchange in S4714 for DM--proton scattering, evaluated at the current experimental upper bounds on $\sigma_{\chi p}$ as a function of DM mass, for spiked (pink) and NFW (green) density profiles. The dotted horizontal line marks $|\mathcal{L}_{\mathrm{DM}}| = L_\star$. The bands indicate the effect of varying $\gamma$ from 0.5 to 1.5. The vertical dotted line separates the region where LZ cosmic-ray-boosted DM constraints set the cross section from the region where standard direct-detection measurements apply.}
\label{fig:p_implied}
\end{figure}

Figure~\ref{fig:p_implied} shows the implied DM-induced luminosity when $\sigma_{\chi p}$ is set to current direct detection upper bounds. For the spiked profile, the luminosity meets or exceeds $L_\star$ across a wide mass range. DM scattering at cross sections consistent with present bounds can provide a luminosity comparable to S4714's. Even for an NFW profile with $\gamma=1.5$, the implied luminosity is well above $L_\star$ for a range of masses. For a more modest slope ($\gamma = 1$), significant energy transfers of order $10\%\,L_\star$ remain achievable within present experimental bounds.
 
\subsection{DM--electron scattering}
 
\begin{figure}[t]
\centering
\includegraphics[width=0.95\columnwidth]{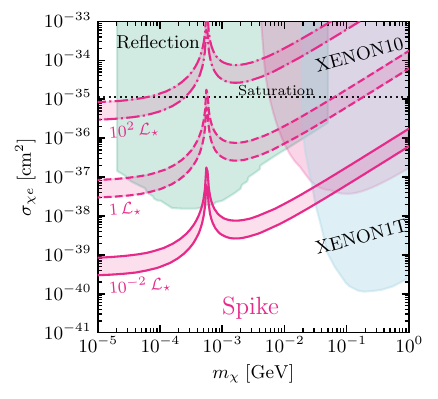}
\caption{Analytic orbit-averaged DM-induced energy exchange in S4714 as a fraction of stellar luminosity, as a function of DM--electron scattering cross section $\sigma_{\chi e}$ and DM mass $m_\chi$, for spiked density profiles, where the band shows the difference between $\gamma=1$ (top) and $\gamma =1.5$ (bottom). Existing experimental constraints are shown from solar reflection~\cite{An:2021qdl}, PandaX~\cite{PandaX:2025rrz}, XENON10~\cite{Essig:2017kqs}, and XENON1T~\cite{XENON:2019gfn}.}
\label{fig:e_bounds}
\end{figure}
 
Figure~\ref{fig:e_bounds} shows the analytic orbit-averaged energy exchange for DM--electron scattering, together with current experimental constraints. For the spiked profile, the exchange reaches the stellar luminosity at $\sigma_{\chi e} \sim 5\times 10^{-38}~\mathrm{cm}^2$ for $m_\chi \lesssim 0.01~\mathrm{MeV}$, in a region not yet excluded by direct detection experiments. While the electron-scattering scenario is more constrained by existing experiments than the proton-scattering case, it is particularly sensitive at sub-MeV masses, where the DM kinetic energy is comparable to or below the electron thermal energy.

 

\subsection{Discussion}
\label{sec:discussion}

The MC confirms the analytic approach across most of the $(m_\chi,\,\sigma_{\chi p})$ plane. Near the heating--cooling transition, the mean-scatter treatment cannot resolve individual collisions and local thermalization. The MC tracks each scatter against the local Maxwell--Boltzmann distribution and produces a broader, smoother transition boundary.

At larger cross sections, the heating--cooling transition shifts to lower masses. Strongly interacting DM thermalizes in the star's outer layers, where the lower effective temperature reduces $m_\chi^{\mathrm{crit}}$ (Eq.~(\ref{eq:critical_mass})).

We have shown that viable cross sections can induce kinetic DM heating at luminosities of $\mathcal{O}(L_\star)$ with DM densities accessible to existing stellar populations near Sagittarius~A*, suggesting a new route toward dark-star phases. Unlike annihilation-powered dark stars~\cite{Spolyar:2007qv, Iocco:2008xb, Scott:2008ns, Moskalenko:2006mk, Freese:2015mta}, this elastic-scattering mechanism applies to a wider class of DM models, including asymmetric DM. Whether a stable dark-star phase can be sustained along an eccentric orbit, and what observational signatures it would produce, remain open questions.

The dominant systematic uncertainty is spike survival under stellar relaxation~\cite{Bertone:2005hw, Balaji:2023hmy, Akita:2025dhg, karydas2026survivaldarkmatterspikes}, which could reduce the energy exchange by one to two orders of magnitude. Recent work modeling specific depletion mechanisms finds comparable or larger softening, e.g. order-of-magnitude to $\sim\!60\%$ depletion near S4714's pericenter from repeated close encounters~\cite{Sharpe:2026nqq}, or relaxation toward $\gamma_{\mathrm{sp}}\approx 1.5$ within a few Gyr~\cite{Herrera:2026hal}. Additional uncertainties arise from the large observational uncertainties in S4714's mass ($2.0^{+2.0}_{-1.0}~M_\odot$) and radius, and from our assumption of solar interior conditions for Galactic Center stars. The pericenter distance carries a large observational uncertainty ($r_{\mathrm{peri}} = 12.6 \pm 9.3~\mathrm{AU}$~\cite{2020ApJ...899...50P}). Since $L_\chi \propto r_{\mathrm{peri}}^{-\gamma_{\mathrm{sp}}}$ exactly (from cancellation of $f_{\mathrm{peri}} \propto r_{\mathrm{peri}}^{3/2}$, $v_{\mathrm{peri}} \propto r_{\mathrm{peri}}^{-1/2}$, and $\delta\varepsilon_{\mathrm{peri}} \propto r_{\mathrm{peri}}^{-1}$), the $1\sigma$ pericenter uncertainty propagates directly through the spike index. For $\gamma_{\mathrm{sp}} = 7/3$ it corresponds to factors of $\sim\!3.7$ lower (at $r_{\mathrm{peri}} = 21.9~\mathrm{AU}$) and $\sim\!23$ higher (at $r_{\mathrm{peri}} = 3.3~\mathrm{AU}$) in the predicted exchange. Improved astrometric constraints on S4714's orbit would therefore directly sharpen the cross-section sensitivity. The steep $\gamma = 1.5$ spike profile is additionally in tension with the recent GRAVITY constraint on the enclosed mass within S2's orbit~\cite{GRAVITY:2024tth}, and results for that profile should be regarded as an upper bound.

Future work includes refining stellar models for the Galactic Center environment, including gravitational acceleration during in-fall, extending the analysis to additional S-stars (e.g., S2, S62) to break DM-astrophysics degeneracies, and embedding the energy-exchange rate in a stellar evolution code to derive reliable cross-section constraints. S4716~\cite{2022ApJ...933...49P, GRAVITY:2021xju} (Table~\ref{tab:stars}) is a natural next target. Its larger periapsis fraction, lower DM density at pericenter, and slower velocity yield a predicted orbit-averaged DM luminosity $\sim\!22$ times smaller than S4714 at fixed cross section, but its tighter astrometric constraints make the estimate more reliable. Velocity-dependent cross sections, exothermic DM, and long-range interactions may further enhance sensitivity. Improved astrometry and multi-epoch photometry of short-period S-stars could reveal a DM-induced energy excess, though extracting such a signal requires careful modeling of astrophysical variability. This would provide a new probe of sub-GeV dark matter at the Galactic Center. Finally, other galaxies such as NGC 1068 could host a DM spike even if the Milky Way does not~\cite{Akita:2025dhg}, and the formalism extends directly to those cases.

 

\section{Conclusion}
\label{sec:conclusion}

We have computed the orbit-averaged elastic DM energy exchange with the star S4714, covering spin-independent DM--proton and DM--electron scattering. For a spiked density profile and proton scattering, we find that the exchange reaches $\mathcal{O}(L_\star)$ across a wide range of DM masses, $10^{-2}~\mathrm{MeV} \lesssim m_\chi \lesssim 10~\mathrm{GeV}$, at cross sections not currently excluded by direct searches, $\sigma_{\chi p} \sim 10^{-36}~\mathrm{cm}^2$ for DM--proton. For electron scattering, we find this to hold for masses $\lesssim 0.05\;\mathrm{MeV}$ with $\sigma_{\chi e} \sim 5\times10^{-38}~\mathrm{cm}^2$ for DM--electron scattering. Both values lie within the range predicted by freeze-in scenarios and related models. Elastic DM scattering can therefore provide a luminosity comparable to that of S4714 while remaining consistent with current experimental constraints.

The result suggests an annihilation-free route to dark-star evolution, applicable to asymmetric DM and other models without self-annihilation. The relevant cross sections fall within the FIMP parameter space~\cite{Elor:2021swj, Bhattiprolu:2022sdd}, where hadrophilic mediators yield $\sigma_{\chi p} \sim 10^{-36}~\mathrm{cm}^2$ for MeV--GeV masses, making S4714 a compelling multi-messenger target, complementary to other multi-messenger probes of the DM–nucleon cross section~\cite{Meighen-Berger:2025hrq}.. Embedding this rate in a stellar evolution code and extending the analysis to additional S-stars would yield reliable cross-section limits and break degeneracies between DM density and astrophysical uncertainties.

 

\section*{Acknowledgments}

We are grateful for helpful discussions with Matheus Hostert, Bhupal Dev, Maxim Pospelov, John Beacom, Katie Auchettl and Dibya Chattopadhyay. This work was supported by the Australian Research Council through Discovery Project DP220101727. S.A.M.B. acknowledges support from the Center for Cosmology and AstroParticle Physics (CCAPP) at Ohio State University and the Department of Physics and Astronomy at the University of Iowa. JLN \& NFB are supported by the Australian Research Council through the ARC Centre of Excellence for Dark Matter Particle Physics, CE200100008. This research was undertaken using the Computational Research, Engineering and Technology Environment (CREATE)~\cite{CREATEHPC} at King's College London. This work is based on the ideas and calculations of the authors, plus publicly available information. IMS and AG are supported
by the U.S. Department of Energy under the award number DE-SC0020262. 
SR  was supported by the Fermi National Accelerator Laboratory (Fermilab), a U.S. Department of Energy, Office of Science, HEP User Facility.

\newpage
\appendix
\counterwithin{figure}{section}
\vspace{0.75cm}
\centerline{\Large {\bf Appendices}}
\vspace{0.75cm}

\section{Baryonic Feedback and Spike Survival}
\label{app:feedback}

We estimate the impact of baryonic feedback due to DM-proton scattering on the formation and survival of the central DM spike. To avoid substantial depletion of the spike over the Milky Way lifetime $t_\mathrm{age}$, we require
\begin{equation}
    t_\mathrm{age} \, n_p \, v_p \, \sigma_{\chi p} \lesssim 1,
\end{equation}
where $n_p$ is the proton number density and $v_p$ a characteristic proton velocity.

We adopt $t_\mathrm{age} = 10~\mathrm{Gyr} \simeq 3\times10^{17}~\mathrm{s}$ and $v_p \lesssim 0.1\,c$. Stellar and gas densities in the Galactic Center yield an effective $n_p \sim 10^7$-$10^8~\mathrm{cm^{-3}}$~\cite{Genzel:2010zy,Schodel:2007er,Schoedel:2008ny,Baganoff:2001ju,1996A&ARv...7..289M}.

Combining these estimates yields the approximate cross-section threshold above which baryonic feedback becomes important for spike survival,
\begin{equation}
\label{eq:feedback_xsec}
    \sigma_{\chi p} \lesssim 10^{-35}~\mathrm{cm^2}.
\end{equation}

This estimate assumes the spike forms via an isothermal core model~\cite{BetancourtKamenetskaia:2025ivl}, in line with earlier studies of feedback effects~\cite{Spergel:1999mh, Kaplinghat:2015aga}. The cross sections at which the DM-induced energy exchange approaches the stellar luminosity (see Figures~\ref{fig:p_mc_high} and~\ref{fig:p_implied}) lie near but do not substantially exceed this threshold, indicating that detailed treatment of baryonic feedback would only mildly affect our conclusions.

\section{Dark Matter Velocity Distribution at Periapsis}
\label{app:dm_velocities}

The analytic and Monte Carlo approaches treat DM as effectively at rest in the Galactic frame, so that the DM velocity in the star's rest frame equals the stellar orbital speed $v_\star$. Here we estimate the size of this approximation.

Inside the sphere of influence of Sgr\,A$^*$, the DM velocity distribution is set by the black hole potential $\Psi(r) = GM_\mathrm{BH}/r$ and is obtained by Eddington inversion of the density profile~\cite{Zhang:2024hrq, Zhang:2025mdl}. For a power-law density $\rho\propto r^{-\gamma}$ in a Keplerian potential, the isotropic distribution function is $f(\mathcal{E})\propto\mathcal{E}^{\gamma-3/2}$, and the velocity distribution at radius $r$ is
\begin{equation}
    g(v) \propto v^2\!\left(\frac{v_{\rm esc}^2 - v^2}{2}\right)^{\!\gamma-3/2}, \quad 0\leq v\leq v_{\rm esc},
    \label{eq:gv}
\end{equation}
where $v_{\rm esc}(r) = \sqrt{2GM_\mathrm{BH}/r} = \sqrt{2}\,v_{\rm circ}(r)$.

\paragraph{Gondolo-Silk spike ($\gamma_{\rm sp}=7/3$).} The exponent in Eq.~(\ref{eq:gv}) is $\gamma-3/2 = 5/6$. Using the Beta-function moments of $g(v)$ gives
\begin{equation}
    \langle v^2\rangle = \tfrac{9}{20}\,v_{\rm esc}^2 = \tfrac{9}{10}\,v_{\rm circ}^2,
    \quad v_{\rm rms} \approx 0.95\,v_{\rm circ}.
\end{equation}

\paragraph{NFW profile ($\gamma=1$).} The exponent is $\gamma-3/2=-1/2$, so $g(v)$ diverges as $v\to v_{\rm esc}$, reflecting the dominance of highly eccentric orbits. The moments give
\begin{equation}
    \langle v^2\rangle = \tfrac{3}{4}\,v_{\rm esc}^2 = \tfrac{3}{2}\,v_{\rm circ}^2,
    \quad v_{\rm rms} \approx 1.2\,v_{\rm circ}.
\end{equation}

\paragraph{Correction to $L_\chi$.} Since the DM velocity distribution is isotropic with zero mean, the angle-averaged relative speed satisfies $\langle|v_{\rm rel}|\rangle > v_\star$ for any non-zero DM dispersion. Because $L_\chi\propto v_{\rm rel}^3$ (flux $\propto v_{\rm rel}$, energy per scatter $\propto v_{\rm rel}^2$), the correction factor is
\begin{equation}
    \frac{\langle v_{\rm rel}^3\rangle}{v_\star^3} = 1 + \frac{2\langle v^2\rangle}{v_\star^2} + \frac{\langle v^4\rangle}{5\,v_\star^4},
    \label{eq:vrel_corr}
\end{equation}
where the average is over $g(v)$ and we used $v_\star = \sqrt{2}\,v_{\rm circ}(r_{\rm peri})$ for the high-eccentricity periapsis speed. Evaluating with the moments above gives correction factors of $\approx 2.0$ (spike) and $\approx 2.6$ (NFW), corresponding to shifts of $\sim 0.3$--$0.4$ in the equal-luminosity cross-section contours. The Monte Carlo approach samples DM entry velocities from the Standard Halo Model~\cite{Drukier:1986tm}, whose dispersion $\sigma_{\rm SHM}\sim 220~{\rm km\,s}^{-1}\ll v_\star\approx 10{,}000~{\rm km\,s}^{-1}$ at periapsis similarly underestimates the DM velocity in the spike, so both methods carry the same systematic underestimate.

\section{Stellar Parameters and Numerical Details}
\label{app:details}

In this appendix, we provide the stellar and orbital parameters of the S-stars used in our analysis and verify the optically thin assumption underlying the analytic approach.

Table~\ref{tab:stars} lists the photometric, physical, and orbital parameters for S-stars near Sagittarius~A*. We verify that S4714 is optically thin to DM-proton scattering using a uniform-density approximation. For S4714 with mass $M_\star \simeq 2\,M_\odot$ and radius $R_\star \simeq 2\,R_\odot$, hydrogen mass fraction $X = 0.7$, and $m_p = 1.6726\times10^{-24}~\mathrm{g}$, the stellar volume implies a mean mass density $\rho \simeq 0.35~\mathrm{g~cm^{-3}}$ and proton number density
\begin{equation}
n_p = \frac{X\,\rho}{m_p} \simeq 1.5\times10^{23}~\mathrm{cm^{-3}}.
\end{equation}
For a central transit and a benchmark cross section $\sigma_{\chi p} = 10^{-40}~\mathrm{cm}^2$, the optical depth is $\tau = n_p\,\sigma_{\chi p}\,(2R_\star) \ll 1$. This argument holds for cross sections $\lesssim 10^{-35}~\mathrm{cm}^2$; for larger cross sections, multiple scatterings per transit become important and the semi-analytic treatment underestimates the energy exchange, making those results conservative lower bounds. The MC calculation (Section~\ref{sec:mc} and Figure~\ref{fig:p_mc_high}) provides the accurate treatment in the high-optical-depth regime.

\begin{table*}[t]
\centering
\footnotesize
\begin{tabular}{lcccccccccc}
\hline\hline
Star & Mag & Mass & $r_{\mathrm{p}}$ & $r_{\mathrm{a}}$ & $v_{\mathrm{p}}$ & $\Gamma$ & $z_{\mathrm{gr}}c$ & $\delta\phi$ & $\dot{\Omega}_{\mathrm{LT}}$ \\
 & [mag$_K$] & [$M_\odot$] & [AU] & [AU] & [km\,s$^{-1}$ (\%$c$)] & [$10^{-4}$] & [km\,s$^{-1}$] & [arcmin] & [arcsec\,yr$^{-1}$] \\
\hline
S62    & 16.1 & 6.1                   & $17.8\pm7.4$   & $1462.4\pm11.0$ & $20124\pm4244$\,($6.7\pm1.4$)       & $46$   & $685.5\pm288.6$ & $74.7\pm31.0$   & $5.1\pm3.2$ \\
S4711  & 18.4 & 2.2                   & $143.7\pm18.8$ & $1094.7\pm28.7$ & $6693\pm494$\,($2.2\pm0.2$)         & $5.6$  & $84.5\pm11.8$   & $10.3\pm1.3$    & $0.34\pm0.07$ \\
S4712  & 18.4 & 2.2                   & $2366\pm120$   & $5075\pm122$    & $1449\pm54$\,($0.48\pm0.02$)        & $0.34$ & $5.1\pm0.4$     & $0.81\pm0.05$   & $(5.1\pm0.5)\times10^{-4}$ \\
S4713  & 18.5 & 2.1                   & $1073\pm110$   & $2234\pm144$    & $2141\pm153$\,($0.71\pm0.05$)       & $0.75$ & $11.3\pm1.3$    & $1.8\pm0.1$     & $(5.8\pm1.1)\times10^{-3}$ \\
S4714  & 17.7 & $2.0^{+2.0}_{-1.0}$  & $12.6\pm9.3$   & $1670\pm10$     & $23928\pm8840$\,($8\pm3$)           & $64$   & $966.1\pm713.5$ & $104.6\pm76.3$  & $7.0\pm7.6$ \\
S4715  & 17.8 & 2.8                   & $894\pm83$     & $1480\pm122$    & $2253\pm129$\,($0.75\pm0.04$)       & $0.90$ & $13.6\pm1.4$    & $2.4\pm0.2$     & $(1.4\pm0.4)\times10^{-2}$ \\
S4716  & 17.0 & $4.0$                 & $99\pm8$       & $706\pm11$      & $7956\pm381$\,($2.65\pm0.13$)       & $8.3$  & $124.8\pm10.1$  & $15.4\pm1.0$    & $1.20\pm0.13$ \\
\hline
S2     & 2.75 & $13.6^{+2.2}_{-1.8}$ & $119.3\pm0.3$  & $1949.9\pm2.8$  & $7582\pm8$\,($2.527\pm0.003$)       & $6.8$  & $101.7\pm5.0$   & $11.7\pm0.6$    & $0.19\pm0.02$ \\
\hline\hline
\end{tabular}
\caption{Photometric, physical, and orbital parameters for S-stars near Sagittarius~A*~\cite{2020ApJ...899...50P, 2017ApJ...847..120H, Cai:2018fkc, 2022ApJ...933...49P, GRAVITY:2021xju}. Photometric results for S62 and S4711--S4716 use S2 as a reference; $\Gamma = r_{\mathrm{s}}/r_{\mathrm{p}}$ is the relativistic parameter ($r_{\mathrm{s}}$ the Schwarzschild radius), $z_{\mathrm{gr}}c$ the gravitational redshift, $\delta\phi$ the Schwarzschild precession, and $\dot{\Omega}_{\mathrm{LT}}$ the Lense-Thirring precession rate for spin parameter $a = 0.5$. S4716 parameters from~\cite{2022ApJ...933...49P}. \textit{S4714 has the smallest pericenter distance and highest pericenter velocity of any confirmed S-star, making it the most sensitive probe of the inner DM distribution.}}
\label{tab:stars}
\end{table*}


\section{Periapsis Fraction and Stellar Correction Factors}
\label{app:fperi_deriv}

We derive the periapsis fraction $f_{\rm peri}$ and the stellar correction factors $A$ and $f_T$ entering Eq.~(\ref{eq:lum_fa}), and compare two independent methods for computing $f_{\rm peri}$.

\paragraph{Keplerian derivation of $f_{\rm peri}$.}
We estimate the fraction of the orbital period spent near periapsis via a parabolic approximation. At periapsis the radial velocity vanishes; the outward radial acceleration for an orbit with eccentricity $e$ is $\ddot{r}|_{r_p} = \mu e/r_p^2$, where $\mu = GM_{\mathrm{BH}}$, so the near-periapsis motion is $r(t) \approx r_p + \frac{1}{2}(\mu e/r_p^2)t^2$. We define the periapsis region as $r \leq r_p(1 + f_r)$, setting $f_r = 2$. This choice captures the dominant contribution to the orbit-averaged DM energy exchange rate: for a spike profile with adiabatic index $\gamma_{\mathrm{sp}}$, the time-averaged integrand $\langle \rho v^3 \rangle$ scales as $r^{-(\gamma_{\mathrm{sp}}+1)}$, and the fraction of total heating within the periapsis region is $F(f_r) \approx 1 - (1+f_r)^{-\gamma_{\mathrm{sp}}}$. For $f_r = 2$ and $\gamma = 1$ ($\gamma_{\mathrm{sp}} \approx 2.33$), $F \geq 90\%$; for an NFW profile with $\gamma = 1$ it captures $\sim 66\%$.

The eccentric anomaly $e_{\mathrm{max}}$ at the boundary of the periapsis region is defined by $r_p(1+f_r) = a(1 - e\cos e_{\mathrm{max}})$, giving $\cos e_{\mathrm{max}} = 1 - f_r(1-e)/e$; near periapsis this simplifies to $e_{\mathrm{max}} \approx \sqrt{2f_r(1-e)/e} \ll 1$. The corresponding mean anomaly is $M_{\mathrm{max}} = e_{\mathrm{max}} - e\sin e_{\mathrm{max}}$, and the fraction of the full period spent within the periapsis region follows from orbital symmetry as $f_{\mathrm{peri}} = M_{\mathrm{max}}/\pi$. Taylor-expanding $M_{\mathrm{max}}$ in $e_{\mathrm{max}}$ and retaining the next-to-leading correction gives
\begin{equation}
f_{\mathrm{peri}} = \frac{(1-e)^{3/2}}{\pi}\sqrt{\frac{2f_r}{e}}
\left(1 + \frac{f_r}{3}\right),
\label{eq:f_peri}
\end{equation}
which agrees with the exact Kepler result to better than $0.5\%$ for S4714 and gives $f_{\rm peri} \simeq 0.002$ for $e = 0.985$.

\paragraph{Orbital integral method.}
An independent estimate follows from integrating the luminosity integrand over the full orbit. Away from the heating/cooling transition, the luminosity scales as $L_\chi \sim \rho(r)\,v^\beta$, where $\beta = 3$ for the heating regime and $\beta = 1$ for cooling, with $\rho(r) \sim r^{-\gamma_{\rm sp}}$. Expressing radius, velocity, and the time element in terms of orbital angle $\phi$,
\begin{equation}
    r(\phi) = \frac{a(1-e^2)}{1+e\cos\phi},
\end{equation}
\begin{equation}
    v(\phi) = \left[\frac{GM(1 + 2e\cos\phi + e^2)}{a(1-e^2)}\right]^{1/2},
\end{equation}
and
\begin{equation}
    \frac{dt}{d\phi} = \frac{(1-e^2)^{3/2}}{n(1+e\cos\phi)^2} ,
\end{equation}
with $n = 2\pi/T_{\rm period}$. We define the angular luminosity $\mathcal{L}(\phi) = L\,dt/d\phi$:
\begin{equation}
    \mathcal{L}(\phi) = \mathcal{C}\,\frac{(1 + e^2 + 2e\cos\phi)^{\beta/2}}{(1+e\cos\phi)^{2-\gamma_{\rm sp}}},
\end{equation}
where $\mathcal{C}$ collects all $\phi$-independent factors. The periapsis fraction then follows from normalizing the full orbit integral by the periapsis value,
\begin{equation}
    f_{\rm peri} \simeq \frac{(1-e^2)^{3/2}}{2\pi(1+e)^{\gamma_{\rm sp}+\beta}}
    \int_0^{2\pi} \frac{(1+e^2+2e\cos\phi)^{\beta/2}}{(1+e\cos\phi)^{2-\gamma_{\rm sp}}}\,d\phi.
\end{equation}
This approach requires no free radial parameter but depends on the regime ($\beta$) and spike index ($\gamma_{\rm sp}$). For S4714 ($e = 0.985$) and $\gamma_{\rm sp} \in (1,\,7/3)$, the integral gives $f_{\rm peri} \sim (0.5\text{--}3)\times10^{-3}$, consistent with the Keplerian estimate of Eq.~(\ref{eq:f_peri}).

The two methods for computing $f_{\rm peri}$ differ in both their conceptual basis and their inputs. The Keplerian approach treats $f_{\rm peri}$ as a geometric quantity, the fraction of the orbital \textit{period} spent within the radial shell $r \leq r_p(1+f_r)$, and is independent of the luminosity scaling $\beta$ and spike index $\gamma_{\rm sp}$. This comes at the cost of the free parameter $f_r$, calibrated to $f_r = 2$ by requiring the periapsis region to capture $\gtrsim 90\%$ of the orbit-integrated exchange for a spike profile. The orbital integral method instead computes $f_{\rm peri}$ as a luminosity-weighted quantity, the fraction of the total orbit-integrated \textit{luminosity} concentrated near periapsis, with no free parameter, but requires specifying $\beta$ and $\gamma_{\rm sp}$. The two quantities coincide only if the luminosity is uniform within the periapsis region. In general, because both $\rho$ and $v$ vary along the orbit, the luminosity-weighted $f_{\rm peri}$ will differ from the time-fraction by a factor that depends on $\gamma_{\rm sp}$ and $\beta$. For steep spikes with $\beta = 3$, the integrand is more sharply peaked near periapsis than the time element alone, so the orbital integral method tends to give a somewhat smaller $f_{\rm peri}$ than the Keplerian estimate. Still, the two agree within a factor of a few across the relevant parameter space. We use Eq.~(\ref{eq:f_peri}) throughout the main text because it is closed-form and regime-independent.

\paragraph{Stellar correction factors.}
The factors $A$ and $f_T$ in Eq.~(\ref{eq:lum_fa}) arise from replacing the uniform-density and uniform-temperature assumptions underlying Eq.~(\ref{eq:lum_peri}) with chord-averaged quantities evaluated on the actual stellar profiles.

For the column correction, Eq.~(\ref{eq:lum_peri}) uses the scattering probability $P_{\rm scat} = n_{\rm core}R_\star\sigma_{\chi p}$, which assumes uniform density $n_{\rm core}$ along the full stellar diameter. For the actual radial profile $n_p(r)$, a DM particle at impact parameter $b$ accumulates a column $\int_{-\ell}^{\ell}n_p(\sqrt{b^2+x^2})\,dx$ with $\ell=\sqrt{R_\star^2-b^2}$. Averaging over impact parameters weighted by the incident DM flux ($\propto 2\pi b\,db$) gives the mean column
\begin{equation}
    \Sigma_n = \frac{2}{R_\star^2}\int_0^{R_\star}b\int_{-\ell}^{\ell}n_p\!\left(\!\sqrt{b^2+x^2}\right)dx\,db,
    \label{eq:sigma_n}
\end{equation}
and $A = \Sigma_n/(n_{\rm core}R_\star)$ is the ratio of the actual chord-averaged column to the uniform estimate. Since $n_p(r)$ peaks toward the stellar core, $A < 1$ in general.

The temperature correction proceeds analogously. The net energy transfer $\delta\varepsilon$ depends on the effective temperature through $T_{\rm eff}$ in Eq.~(\ref{eq:dE}). Weighting the chord integral by $n_p T$ rather than $n_p$ gives the density-weighted mean temperature,
\begin{equation}
    \Sigma_{nT} = \frac{2}{R_\star^2}\int_0^{R_\star}b\int_{-\ell}^{\ell}n_p\!\left(\!\sqrt{b^2+x^2}\right)T\!\left(\!\sqrt{b^2+x^2}\right)dx\,db,
    \label{eq:sigma_nT}
\end{equation}
so that $f_T = \Sigma_{nT}/(\Sigma_n T_{\rm core})$ and $T_{\rm eff} = f_T T_{\rm core}$. Using the profiles $T(r)$, $n_p(r)$, and $R_\star$ from a $2\,M_\odot$ MESA stellar model~\cite{2011ApJS..192....3P,2013ApJS..208....4P,2015ApJS..220...15P,Paxton:2017eie,2019ApJS..243...10P,MESA:2022zpy} we find $A \sim 0.14$ and $f_T \sim 0.4$.

\section{Monte Carlo Simulation Details}
\label{app:mc_details}

Here we describe the MC procedure used to compute the mean net energy deposited per DM transit, $\langle\Delta E\rangle(m_\chi, \sigma_{\chi p}, v_{\mathrm{entry}})$.

For each transit, an impact parameter is drawn uniformly in area, $b = R_\star\sqrt{\xi}$ with $\xi \sim \mathcal{U}[0,1]$, fixing a chord of half-length $\ell = \sqrt{R_\star^2 - b^2}$ and radial coordinate $r(x) = \sqrt{b^2+x^2}$ parameterized by position $x\in[-\ell,\ell]$ along the chord. The DM propagates along the chord accumulating optical depth $\mathrm{d}\tau = n_H(r)\,\sigma_{\chi p}\,\mathrm{d}x$. Here $n_H$ is the local number density of Hydrogen. The position of the next scatter is sampled by drawing $\tau_{\mathrm{next}} = -\ln\xi'$. At each scatter, the target proton velocity is drawn from the local Maxwell-Boltzmann distribution, $\mathbf{v}_p \sim \mathcal{N}(0,\sqrt{k_BT(r)/m_p}\,)^{\otimes3}$, and the post-scatter DM velocity is obtained by boosting to the center-of-mass frame, applying an isotropic random rotation, and boosting back. If the DM speed falls below the local stellar escape velocity, which at its maximum is $v_{\mathrm{esc}} = \sqrt{2GM_\star(r)/R_r}$, the particle is gravitationally captured. Where the escape velocity is calculated at the current position. For a core temperature $T_c$, DM with $m_\chi < m_{\mathrm{evap}} \equiv 2k_B T_c/\beta_{\mathrm{esc}}^2$, where $\beta_{\mathrm{esc}} \equiv v_{\mathrm{esc}}/v_{\mathrm{th}}$ and $v_{\mathrm{th}} = \sqrt{2k_BT_c/m_\chi}$, subsequently thermalizes and evaporates, depositing $\Delta E = E_{\mathrm{entry}} - k_BT_c$, while heavier captured DM deposits $\Delta E = E_{\mathrm{entry}}$. The net energy per escaping transit is $\Delta E = E_{\mathrm{entry}} - E_{\mathrm{exit}}$.

For numerical efficiency, the calculation employs a three-step approach. For $\mu^2 \equiv m_\chi m_p/(m_\chi+m_p)^2 < 0.05$ (i.e., $m_\chi \lesssim 50~\mathrm{MeV}$), the mean deposition is computed deterministically by integrating the Fokker-Planck drift equation along each chord and averaging over impact parameters, giving the exact mean with no statistical noise. When the on-axis optical depth $\tau_{\mathrm{axis}} = 2\sigma_{\chi p}\int_0^{R_\star} n_H\,\mathrm{d}r < 0.1$, an exact analytical formula for the thin-limit single-scatter mean is used. Otherwise, full particle transport is performed with adaptive early stopping once the relative standard error of the mean falls below 1\%.

We pre-compute the table $\langle\Delta E\rangle$ on a grid of $40\times50\times15$ points in $(\log_{10}m_\chi,\,\log_{10}\sigma_{\chi p},\,\log_{10}v_{\mathrm{entry}})$ spanning $m_\chi\in[10^{-5},10]~\mathrm{GeV}$, $\sigma_{\chi p}\in[10^{-50},10^{-25}]~\mathrm{cm^2}$, and $v_{\mathrm{entry}}\in[10^6,10^{9.5}]~\mathrm{cm\,s^{-1}}$, using 1000 particles per grid point. Sampling of orbit locations and velocities  (using the Standard Halo Model~\cite{Drukier:1986tm}) is done evenly across the orbit and iteratively. Multiple simulations are run with increasing sample sizes, until a convergence criterion is reached. Here, we require that the final energy exchange not change by more than 1\% when the sample size is increased by an order of magnitude. We reach this point at $\sim 50000$ evenly distributed samples. 

\clearpage
\bibliography{bibliography}


\end{document}